\documentclass[sigconf,9pt]{acmart}
\usepackage[T1]{fontenc}
\usepackage{graphicx}
\usepackage{makecell,booktabs} 
\usepackage{amsmath} 
\usepackage{url}
\usepackage{comment}
\usepackage{balance}
\usepackage[nameinlink,capitalise]{cleveref}
\usepackage{color}

\definecolor{gg}{RGB}{0, 129, 35}
\definecolor{orangeShallow}{RGB}{255,190,0}

\usepackage[utf8]{inputenc}
\usepackage{tabulary}
\usepackage{colortbl}
\usepackage{xcolor}
\usepackage{hyperref} 

\usepackage[linesnumbered,ruled]{algorithm2e}
\usepackage{setspace} 

\SetAlgoSkip{smallskip}            
\SetAlgoInsideSkip{smallskip}      
\SetAlCapSkip{0ex}                 
\SetAlgoNlRelativeSize{-1}         
\DontPrintSemicolon                
\SetKw{Continue}{continue}
\SetKwInput{KwInput}{Input}                
\SetKwInput{KwOutput}{Output}              
\SetKwComment{Comment}{$\triangleright$\ }{}

\usepackage{longtable, array, booktabs}
\usepackage{multirow}
\usepackage{adjustbox}
\usepackage{tikz}
\usepackage{threeparttable}

\newcommand{\hongce}[1]{\textcolor{red}{[\textbf{hongce}: #1]}}

\copyrightyear{2026}
\acmYear{2026}
\setcopyright{cc}
\setcctype{by}
\acmConference[ICCAD '26]{IEEE/ACM International Conference on Computer-Aided Design}{November 08--12, 2026}{San Jose, CA, USA}
\acmBooktitle{IEEE/ACM International Conference on Computer-Aided Design (ICCAD '26), November 08--12, 2026, San Jose, CA, USA}
\acmDOI{10.1145/3831252.3834042}
\acmISBN{979-8-4007-2873-0/2026/11}

\begin{document}

 \title{NeuroAbs: A Neuro-Symbolic RTL Abstraction Framework for Property Checking Acceleration}

\author{Zhiyuan Yan\textsuperscript{1},
  Xiaofeng Zhou\textsuperscript{2},
  Ziyue Zheng\textsuperscript{1},
  Ziyi Yang\textsuperscript{1},
  Wenbin Che\textsuperscript{1},\\
  Wei Zhang\textsuperscript{2},
  Yangdi Lyu\textsuperscript{1}, and
  Hongce Zhang\textsuperscript{1,2,*}}
\affiliation{%
  \institution{\textsuperscript{1}The Hong Kong University of Science and Technology (Guangzhou)\\
  \textsuperscript{2}The Hong Kong University of Science and Technology\\
  \textsuperscript{*}Corresponding author:
  \href{mailto:hongcezh@hkust-gz.edu.cn}{hongcezh@hkust-gz.edu.cn}}
  \country{}}

\renewcommand{\authors}{Zhiyuan Yan, Xiaofeng Zhou, Ziyue Zheng, Ziyi Yang,
  Wenbin Che, Wei Zhang, Yangdi Lyu, and Hongce Zhang}
\renewcommand{\shortauthors}{Yan et al.}

\begin{abstract}

Formal verification is a crucial technique for ensuring the functional correctness of hardware designs.
In the context of property checking, a key challenge is how to efficiently prove a user-specified property in the face of increasingly complex RTL designs. 
To address this challenge, abstraction techniques are often employed to reduce system complexity and accelerate the verification process. However, prior RTL abstraction methods either require significant manual effort or rely on rule-based techniques that lack flexibility.
This paper introduces NeuroAbs, a neuro-symbolic framework for RTL abstraction. NeuroAbs first uses LLM-assisted RTL analysis to identify signals suitable for abstraction. It then combines LLM-based abstraction with an AST-based symbolic RTL representation to better align the generated abstraction with the intended transformation. The soundness of each abstraction is checked using satisfiability modulo theories (SMT) solving. If the abstraction is too coarse for a successful proof, NeuroAbs applies counterexample-guided abstraction refinement (CEGAR) to iteratively refine the model. Experimental results show that NeuroAbs significantly improves the efficiency of hardware property checking across a range of verification tasks. \looseness = -1
\end{abstract}

\maketitle

\section{Introduction}\label{sec::intro}
Formal verification plays a crucial role in ensuring the correctness of digital circuit designs, particularly through property checking (also known as model checking). It systematically determines whether a given register-transfer-level (RTL) hardware model complies with a specified property. 
Upon  property failure, a counterexample trace will be generated to illustrate how the property is violated by the given RTL design.
Otherwise, the verification engine will construct a formal proof establishing the property's validity. 
Over the years, several model checking algorithms have been proposed for hardware formal verification, including bounded model checking (BMC)~\cite{biere1999symbolic}, k-induction~\cite{sheeran2000checking}, and the property-directed reachability (PDR) algorithm~\cite{een2011efficient}. \looseness = -1



As circuits grow in size and complexity, a major challenge in formal verification is how to efficiently determine the correctness of a given property. To address this, prior research has explored the use of abstraction techniques, which replace the concrete circuit model with a simplified representation that preserves only the essential behaviors relevant to verification. These techniques can be broadly classified into two categories: (1) manual abstraction, and (2) rule-based automatic abstraction. Manual abstraction largely relies on human insight into both the design and the target property. For instance, counter abstraction has been proposed to reduce the complexity of verifying systems that involve counters~\cite{ip2008managing}. Similarly, Hsieh and Levitan~\cite{hsieh1998model} suggested abstracting digital circuit models using structural analysis and relational algebra. Bryant \textit{et al.}~\cite{bryant2002modeling} introduced the CLU (Counter Arithmetic with Lambda Expressions and Uninterpreted Functions) framework to model complex components such as infinite memories. \looseness =-1

Rule-based abstraction, on the other hand, is an automated approach that systematically performs the abstraction process by applying predefined rules or algorithms. For example, Mishchenko \textit{et al.} and Fan \textit{et al.} utilized gate-level abstraction to accelerate the BMC and PDR algorithms, respectively~\cite{mishchenko2013gla,fan2016automatic}. PDR-WLA abstracts the model by replacing signals generated by multipliers, adders,
etc. with primary inputs~\cite{ho2017property}, while IC3IA implements implicit predicate abstraction to 
perform the PDR algorithm at a higher level~\cite{cimatti2014ic3}. Furthermore, AVR~\cite{goel2020avr}, as the champion of the 2020’s hardware model checking competition (HWMCC)~\cite{preiner2020hardware}, employed the syntax-guided abstraction, which implicitly builds the abstraction model by extracting syntactic terms directly from the word-level syntax of the given hardware description. In addition, AVR also performs datapath abstraction by replacing
datapath operations with uninterpreted functions (UFs) to abstract complex arithmetic operations.
Fan \textit{et al.} proposed the datapath propagation technique to generate a more accurate abstract model in AVR for the datapath~\cite{fan2024leveraging}. 
These techniques reduce the complexity of the digital model, enabling a more efficient verification process.\looseness=-1


\begin{figure}[t]
    \centering

    \includegraphics[width=0.27\textwidth]{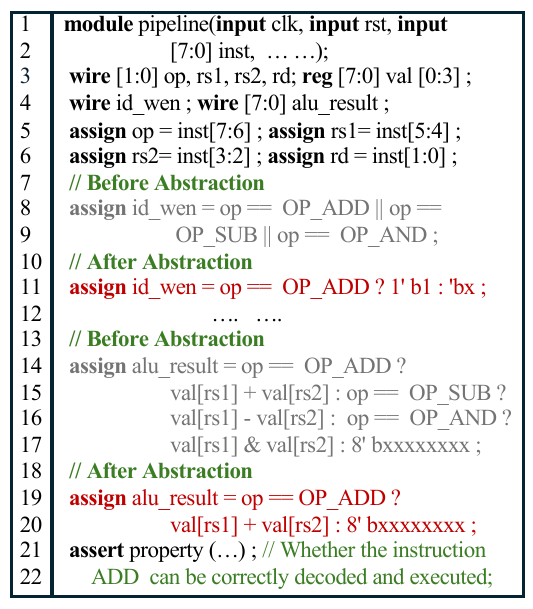}
            \vspace{-3mm}
            \caption{An example of abstraction based on the high-level information of the given property. In this example, \texttt{OP\_ADD}, \texttt{OP\_AND} and \texttt{OP\_SUB} are three different parameters, where $\texttt{OP\_ADD} = 2'b01$, $\texttt{OP\_SUB} = 2'b10$, and $\texttt{OP\_AND} = 2'b11$.}
    \label{fig:motivating example}
\vspace{-5mm}
\end{figure}


In general, however, how to apply proper abstraction is both design-specific and property-specific.
For example, \cref{fig:motivating example} shows a Verilog code snippet of the decoding and ALU logic in a microprocessor.
In the formal verification of this design, the engineer may, for example, use assertions to check the correct RTL implementations for each type of instructions, which is the case in ISA-formal~\cite{huang2018instruction}. It is not hard to see that, when verifying the property for one instruction type (for example, the ADD instruction), the RTL logic related to other instructions (e.g., bitwise AND and subtraction operations) can be safely abstracted. Specifically in this example, the assignment to the signal \texttt{id\_wen} and \texttt{alu\_result} can be simplified by replacing (part of) a concrete expression to a free value (which may also be referred to as unknown value or X-value), since the execution of these statements will not contribute to the outcome of the specified property. 
However, this kind of abstraction relies on a high-level understanding of the RTL design and the property specification.
Traditional rule-based abstraction methods, such as structural abstraction and syntax-guided abstraction, usually are not ``smart'' enough to identify such opportunities per each verification setting, as they rely on predefined terms or syntactic rules rather than semantic reasoning about the property’s intent. Consequently, they fail to recognize abstraction opportunities that depend on functional understanding. More importantly, the abstraction opportunities in our setting depend jointly on the verification scenario, the target property, and the local RTL context, making them difficult to enumerate exhaustively with a fixed set of hand-written rules. On the other hand, manual abstraction construction remains tedious and labor‑intensive.\looseness=-1

\begin{figure*}[t]
    \centering

    \includegraphics[width=0.86\textwidth]{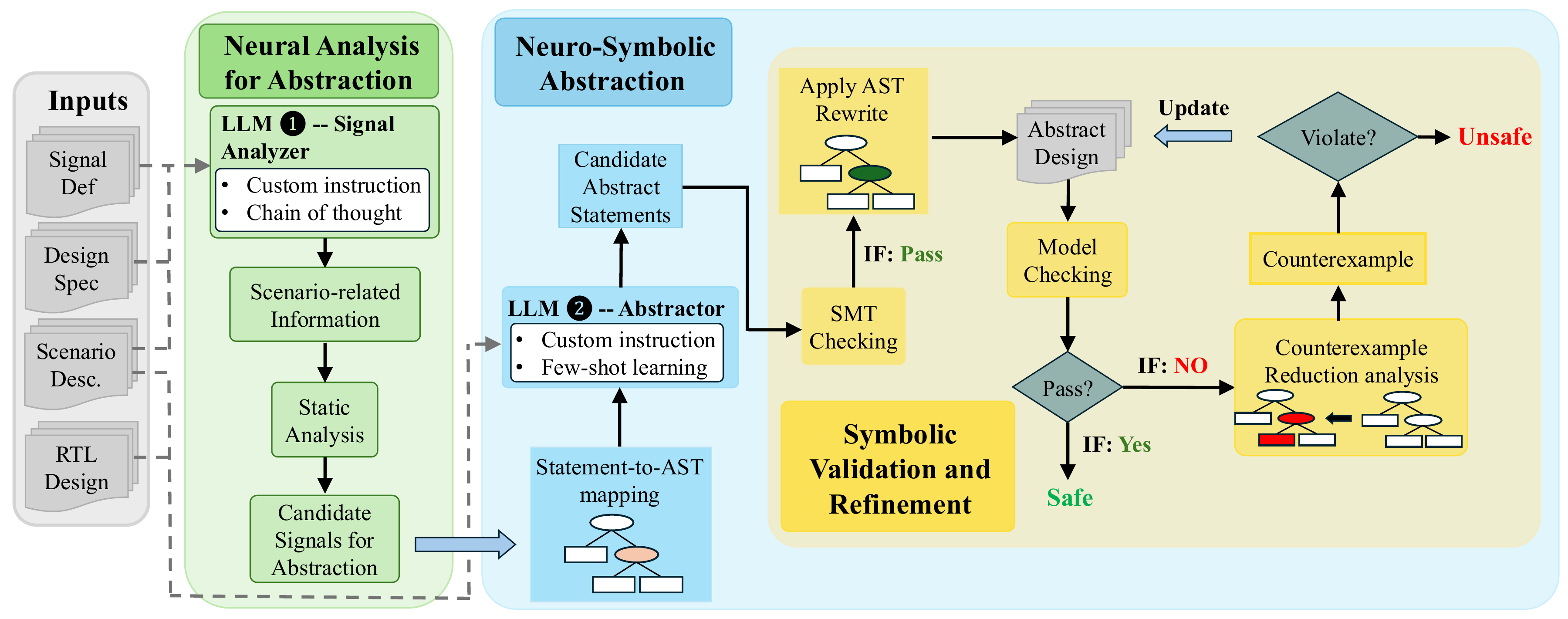}
            \vspace{-3mm}
            \caption{An overview of NeuroAbs.}
    \label{fig:overflow}
\vspace{-3mm}
\end{figure*}

To address the above challenge, we propose NeuroAbs, a neuro-symbolic framework for automated RTL abstraction. While LLMs are effective at generating candidate rewritings from design and property intent, a critical challenge is whether the generated abstraction remains aligned with the intended abstraction transformation. In particular, free-form rewriting lacks explicit structural constraints, and may therefore abstract logic beyond what is justified by the target property, modify logic outside the intended abstraction region, introduce syntactic and semantic errors, or violate the required over-approximation relation. NeuroAbs addresses these issues by combining LLM-based generation with an AST-based symbolic representation of RTL and formal checking. Specifically, NeuroAbs constructs the abstract model in three steps. First, it performs LLM-assisted analysis to identify the signals that are likely relevant to the target property. Second, it traverses the symbolic RTL representation to localize the corresponding statement-level AST nodes and uses another LLM to rewrite the selected RTL only within these designated regions. Third, it uses SMT-based soundness checking and counterexample-guided abstraction refinement (CEGAR)~\cite{clarke2000counterexample} to ensure that the generated abstraction is a sound over-approximation and to refine it when it is too coarse. We evaluate NeuroAbs on several representative verification tasks, including those from RISC-V Formal~\cite{riscvformal}, instruction-level abstraction refinement checking~\cite{huang2018instruction,fang2023r}, and the verification of an I2C peripheral. Experimental results show that NeuroAbs accelerates formal proof search and significantly improves bug-finding efficiency with BMC.

Overall, this paper makes the following contributions:
\begin{itemize}
\item We propose NeuroAbs, a neuro-symbolic framework for RTL abstraction. It uses one LLM to analyze design and property intent and identify abstraction candidates, and another LLM to generate the abstraction. To the best of our knowledge, it is the first work to apply LLMs to RTL abstraction.
    \item We introduce a neuro-symbolic abstraction flow in which the neural component proposes scenario-aware abstraction candidates from the design and property context, while the symbolic component constrains rewriting to AST-level targets, validates the generated abstractions through SMT checking, and iteratively refines the abstract model through a customized CEGAR procedure when the abstraction is too coarse.\looseness = -1
    \item We evaluate NeuroAbs on a range of model checking problems and show that it improves verification efficiency across different tasks and algorithms.      
\end{itemize}




\section{Background}

\label{sec::background}
\subsection{Hardware Model Checking and Abstraction}
\label{sec::background_abstraction}
 An RTL model can be represented as a finite state transition system: $M := \langle V,  \mathit{Init}(V), \mathit{Tr}( V, V') \rangle$. Here, $V$ represents state variables, and $V'$ indicates next-state variables. $\mathit{Init}(V)$ specifies the initial states, and $\mathit{Tr}(V, V')$ defines the transition relation. The input variables in RTL are treated as free state variables which are not bounded by the transition relation, and therefore, we omit them in the formulation. \looseness = -1


Given a safety property $P(V)$, the hardware model checking algorithm determines whether $P(V)$ holds for all states reachable from $\mathit{Init}(V)$.
If $P(V)$ is valid for all reachable states, the property is safe, and a formal proof can be constructed.
Otherwise, the system is regarded as unsafe, and a counterexample trace is produced.

To improve scalability, abstraction is commonly used to construct a model that focuses on the essential behaviors of the RTL while simplifying certain details. The abstract transition system is formally defined as  $\hat{M} := \langle \hat{V},  \hat{\mathit{Init}}(V), \hat{\mathit{Tr}}( V, V') \rangle$, where $\hat{M}$ is the overapproximation of the original system  $M$. The abstraction guarantees that $\mathit{Init} \models \hat{\mathit{Init}}$ and $\mathit{Tr} \models \hat{\mathit{Tr}}$, ensuring that the state space of $\hat{M}$ is strictly a superset of the state space of $M$. If the safety property $P(V)$ is proved as safe under the abstract model $\hat{M}$, it also implies that the $M$ is safe. However, if $P(V)$ is unsafe under the $\hat{M}$, the generated counterexample may be spurious due to the too coarse abstraction. Therefore, the abstract model must be refined using the CEGAR method to eliminate this counterexample. As there could be more than one spurious counterexample, the process of abstraction and refinement may form a loop, commonly referred to as the CEGAR loop. \looseness = -1

\subsection{Scenario-based Verification Setup}


Modern hardware often supports multiple configurations or operating modes, each exhibiting distinct functional behaviors. These variations are typically verified separately, with each case requiring a customized setup. We refer to each such focused verification effort as a verification scenario, which defines the operational context,  relevant assumptions, and the properties to be formally checked (i.e., property specification). For example, in processor verification, one scenario may target the correct execution of one instruction class (e.g., ALU instructions with immediates), while another may check consecution of program counters. Similarly, for a Serial Peripheral Interface (SPI) controller, verification scenarios may cover initialization, data transmission, and data reception, etc. \looseness=-1


Given a verification scenario, not all part of the RTL model is exercised. As shown in \cref{fig:motivating example}, when verifying the ADD instruction, logic related to other instructions can be safely abstracted. However, conventional techniques including the Cone of Influence may miss such opportunities because they rely on structural rather than functional analysis. Consequently, we propose using LLMs to identify the logic that is truly relevant to each verification scenario. \looseness = -1

\section{Our Method}
\label{sec::method}

\subsection{Workflow Overview}
\cref{fig:overflow} illustrates the abstraction process in NeuroAbs. Rather than allowing the LLM to directly rewrite RTL in a free-form manner, NeuroAbs follows a neuro-symbolic flow in which the neural component proposes scenario-aware abstraction candidates and the symbolic component constrains, validates, and refines them. NeuroAbs proceeds in three stages. First, the RTL Signal Analyzer takes as input the design specification, the verification scenario, and signal definitions, and identifies signals that are likely relevant to the target property. Second, starting from the selected signals, NeuroAbs locates the corresponding RTL statements, maps them to AST statement nodes, and asks the Abstractor to generate candidate abstractions only at the selected rewrite points. These candidate rewrites are then checked by SMT, and only those that satisfy the required over-approximation are applied to update the abstract design. Third, the resulting abstract design is verified against the target property. If the abstraction is still too coarse and introduces spurious counterexamples, NeuroAbs uses a customized CEGAR loop with counterexample reduction to refine the abstract model. \looseness=-1

\begin{figure}[t]
    \centering

    \includegraphics[width=0.9\linewidth]{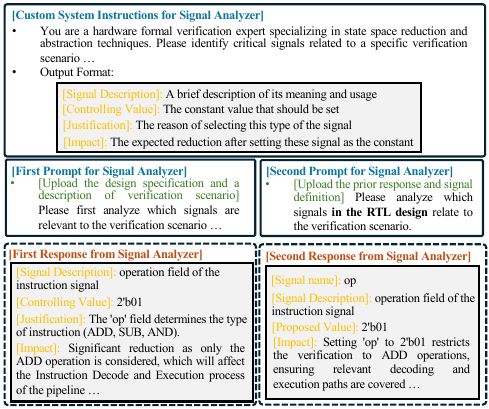}
    \vspace{-3mm}
            \caption{The custom instruction, COT prompt and response of the Signal Analyzer.}
    \label{fig:LLM1}
\vspace{-5mm}
\end{figure}


\subsection{Candidate Signal Identification for Abstraction}\label{sec::rtl-code-region-identification}

As RTL can be very lengthy, it is paramount to first narrow down where abstraction may be applied. The goal of this stage is to identify candidate signals for abstraction, rather than determine the final rewrite itself. A naive approach would be to provide the entire RTL code to an LLM for analysis. However, given the limited context length of LLMs, directly feeding thousands of lines of RTL code for abstraction is infeasible.


To address this problem, we propose a hybrid approach that integrates an LLM with RTL static analysis to identify abstraction candidates. The LLM analyzes the design specification and verification scenario description, along with formal properties and environment setup when available, to extract relevant information in natural language that traditional rule-based methods cannot easily capture. It then suggests signals related to the scenario, referred to as \textit{scenario-related signals}. Starting from these signals, RTL static analysis is performed to further track the corresponding RTL logic where abstraction may later be applied. At this stage, the output is a localized set of candidate signals for abstraction, rather than a concrete rewrite. 
Inspired by chain-of-thought (COT)~\cite{wei2022chain}, we build an LLM-based pipeline, called the Signal Analyzer, to generate the \textit{scenario-related signals} in two steps. First, it interprets signal behaviors based on the design specifications 
and the scenario descriptions. Then, it outputs specific signal names according to the prior analysis and the signal definitions in the RTL model. \cref{fig:LLM1} shows a brief example of the Signal Analyzer’s prompts and responses. This LLM is guided by customized system instructions containing background information of RTL verification, the general setting of NeuroAbs, and the subsequent verification steps. 
We also provide a unified template for the initial response, which directs the LLM to describe each signal’s meaning and controllable value within the verification scenario. The template also requires the LLM to justify its signal selection and explain the expected 
effect of simplifying the related logic, enabling engineers to better understand the LLM’s choices.

After generating the initial response, the next step of the  Signal Analyzer is to map the scenario-related signals in the documents to their actual names in RTL. We provide the LLM with the signal definitions in the RTL module together with its first response.
As shown in \cref{fig:LLM1}, the analyzer aligns the signal declaration with the generated descriptions, and outputs the response accordingly.



Given the initial set of scenario-related signals, we apply RTL static analysis to identify the corresponding code regions relevant to these signals, such as statements connected through control or data dependency. In some verification scenarios, certain control signals are assigned or assumed to hold fixed values. The analysis can also identify opportunities of simplification due to fixed values through techniques such as constant propagation~\cite{wegman1991constant}. Note that these fixed values are used only in this static-analysis stage to guide localization. Once the abstraction targets are identified, they are not carried forward as additional assumptions in the generated abstract model or the downstream verification procedure. \looseness = -1


\begin{figure}[!t]
    \centering

    \includegraphics[width=0.27\textwidth]{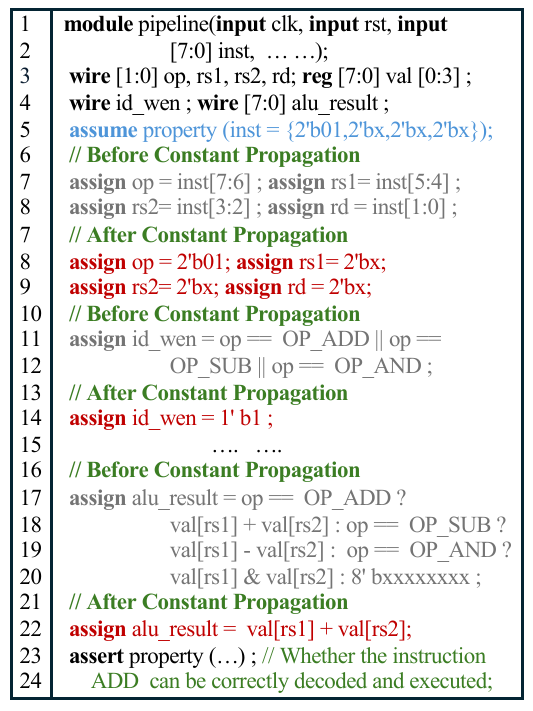}
    \vspace{-3mm}
            \caption{The example of constant propagation.  After we assign a constant value to \texttt{inst}, the signal \texttt{op} is set to \texttt{OP\_ADD} ($2'b01$). Following the constant propagation, \texttt{id\_wen} and \texttt{alu\_result} will evaluate to $1'b1$ and $\texttt{val[rs1]} + \texttt{val[rs2]}$.}
    \label{fig:constant_prop_example}
    \vspace{-3mm}
\end{figure}

\cref{fig:constant_prop_example} presents a simplified example based on the same microprocessor design shown in  \cref{fig:motivating example}. When verifying the ADD instruction, the Signal Analyzer identifies that the opcode field of the instruction word (\texttt{inst}) is a scenario-related signal with a controlled value of $2'b01$, corresponding to the \texttt{OP\_ADD} operation. While the remaining bits indicating source registers and the destination register are not controlled. 
By applying constant propagation, this constant value is propagated to related signals, allowing certain assignments --- such as those to \texttt{id\_wen} and \texttt{alu\_result} --- to be simplified, as illustrated in the example. This simplification suggests that the corresponding statements of these signals are suitable candidates for abstraction, consistent with the earlier discussion in \cref{sec::intro}. This step narrows the abstraction scope to a small set of candidate signals, which are then used to localize statement-level AST nodes in the next subsection.

\subsection{Neuro-Symbolic LLM-based Abstraction}
\label{sec::abstraction}
\begin{figure}[!t]
    \centering

    \includegraphics[width=0.66\linewidth]
    {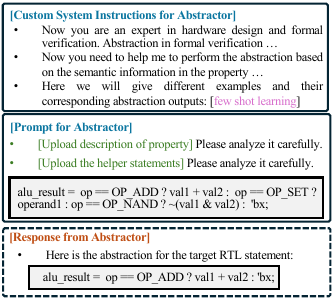}
            \vspace{-3mm}
            \caption{The custom instruction, prompt and response of the Abstractor.}
    \label{fig:LLM2}
    \vspace{-3mm}
\end{figure}

The previous stage has already identified the signals that are relevant to the target property. We next convert RTL into an AST-based symbolic representation and traverse the AST to localize the statement(s) corresponding to these signals. This AST representation provides a precise structural boundary for each candidate statement compared to unconstrained text-level localization in RTL, recovers the enclosing RTL context, and links each rewrite to the subsequent symbolic checking and refinement steps. The Abstractor then generates abstractions only at these selected nodes, and each resulting rewrite is checked to ensure the required over-approximation relation. \looseness = -1
{\bf AST-based localization.}
Let $A$ denote the set of AST nodes in the RTL design, and let:
\[
\Phi_{\mathit{loc}}: A \rightarrow \{\mathit{true}, \mathit{false}\}
\]
be the matching condition induced by the localization result from the previous stage. For a node $a \in A$, $\Phi_{\mathit{loc}}(a)=\mathit{true}$ means that $a$ is a statement-level AST node localized from the selected signals. These nodes serve as the symbolic interface that constrains the neural Abstractor to a finite set of semantically meaningful rewrite targets. We further trace upward to the enclosing constructs, such as ``always'' blocks, ``initial'' blocks, and ``case'' statements, and provide them to the Abstractor as auxiliary structural context.

{\bf Transformation.}
After the target statement is fixed, the second LLM, called Abstractor, performs the transformation. We ask the Abstractor to generate the abstraction on the selected statement-level AST node together with its local structural context. Given this statement and its enclosing structural context, the Abstractor rewrites the corresponding RTL text only within the AST-localized statement region, rather than editing an unconstrained text span. The Abstractor then decides which internal part of the statement should be abstracted according to the property. Formally, let:
\[
\tau: \{ a \in A \mid \Phi_{\mathit{loc}}(a)=\mathit{true} \} \rightarrow \hat{A}
\]
denote the abstraction transformation, where $\tau$ maps each localized statement AST node $a$ to a rewritten AST $\hat{a} \in \hat{A}$. As illustrated in \cref{fig:LLM2}, the Abstractor receives the relevant property specification, the entire target statement, and its enclosing structural context, and rewrites only the part of that statement judged irrelevant to the target property. For example, in the statement computing \texttt{alu\_result}, the addition-related computation is preserved, while the unrelated part can be abstracted with X-values or simplified expressions. In this way, the AST determines \emph{which} statement may be rewritten, while the LLM determines \emph{how} to rewrite the statement. \looseness = -1


Besides the target RTL statement and its context, the Abstractor employs few-shot learning~\cite{wang2020generalizing} by providing a small set of input-output examples (original vs. abstract statements) to increase the probability of generating a correct abstraction. The soundness of each generated abstraction is then checked by an SMT-based rule, as described next. \looseness = -1

\subsection{SMT-based Soundness Constraint}\label{sec::soundness}

Because the LLM-generated abstraction may fail to over-approximate the original RTL design, we impose an additional symbolic constraint using Satisfiability Modulo Theories (SMT)~\cite{barrett2018satisfiability}. For each selected RTL statement, the generated abstraction is accepted only if it satisfies the following SMT checking rule:
 \begin{equation}
\begin{split}
\forall V,\ S(V) \Rightarrow \hat{S}(V)
\end{split}
\end{equation}
where $S(V)$ and $\hat{S}(V)$ denote the SMT expressions induced by the original and abstracted versions of the selected RTL statement, respectively. This rule requires every behavior allowed by the original statement to also be allowed by the abstracted statement, i.e., the abstraction must be an over-approximation.
In many cases, the Abstractor introduces unknown values (X-values) in the abstract statements. Existing SMT solvers do not natively support X-values. To address this issue, we translate each X-value into a fresh variable bound by an existential quantifier, and refine the checking rule as follows: \looseness=-1

 \begin{equation}
 \label{eqa::overapp}
\begin{split}
\forall V,\ \exists Y,\ \left[ S(V, Y) \Rightarrow \hat{S}(V, Y) \right]
\end{split}
\end{equation}
Here, $Y$ denotes the additional variables introduced for the X-values. This formulation lets the SMT solver interpret the X-values as nondeterministic choices and check whether the rewritten statement still over-approximates the original statement. Although quantified bit-vector reasoning is generally harder than quantifier-free solving, the formulas in Equation~\ref{eqa::overapp} remain moderate in size in our setting. Therefore, we do not observe performance issues in soundness checking in our experiments. Since each accepted rewrite over-approximates its original local RTL statement and all unrevised statements remain unchanged, the composed abstract RTL model also remains an over-approximation of the original design. 
\subsection{Customized CEGAR Loop}
\label{sec::cegar}
Once the abstract model is built through the prior steps, the model could be utilized to verify the target property $P(V)$. However, as discussed in \cref{sec::background_abstraction}, abstraction may incur spurious counterexamples if it is too coarse for the property to verify. Consequently, the model needs to be refined to eliminate these spurious counterexamples. Prior research has proposed the counterexample-guided abstraction refinement (CEGAR) paradigm to iteratively block these spurious counterexamples~\cite{clarke2000counterexample}. However, a coarse abstraction often leads to numerous counterexamples, and it may take many iterations to properly refine the model.
It is beneficial if the counterexample could be generalized to better guide the refinement. 
Therefore, we propose a customized CEGAR implementation integrated with counterexample reduction and generalization techniques~\cite{yan2025word} to better pinpoint which rewritten statement(s) actually trigger the violation of the property.
These relevant abstract statements are then reverted to their original form. By focusing on refining abstract statement(s) that trigger a set of counterexample, this approach helps to reduce the number of iterations in the CEGAR loop. We name this implementation as the R-CEGAR method to emphasize the use of counterexample reduction techniques.

\begin{figure}[t]
    \centering

    \includegraphics[width=0.6\linewidth]{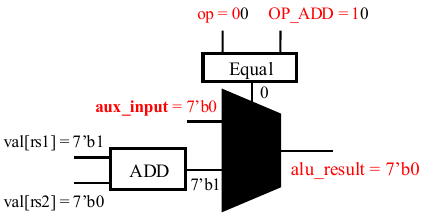}\
    \vspace{-3mm}
            \caption{An example of counterexample reduction for our abstract model. When the target property is violated, assignments can be extracted. The reduced counterexample is highlighted in red.}
    \label{fig:cex_reduction}
    \vspace{-3mm}
\end{figure}

\cref{fig:cex_reduction} illustrates an example of counterexample reduction in our setting.  This example is derived from the abstract statement in \cref{fig:motivating example} (lines 19–20), which computes the value of the signal \texttt{alu\_result}. The statement involves a 2:1 multiplexer controlled by a comparator that outputs 1 during the execution of the addition operation. As mentioned earlier, we conjecture that operations related to subtraction and bitwise AND do not contribute to the property that we want to check. Therefore, we abstract these operations by replacing them with X-values. For each abstraction, we automatically generate an auxiliary input (labeled as \texttt{aux\_input} in the figure) for the X-value. As shown in the counterexample trace, the multiplexer control bit is 0, meaning the result of \texttt{alu\_result} depends solely on the signal \texttt{aux\_input}, without considering the addition of \texttt{val[rs1]} and \texttt{val[rs2]}. As a result, after applying counterexample reduction, it becomes clear that only \texttt{aux\_input} is relevant to the property violation, indicating that the introduced X-value is not a proper abstraction for the properties. This indicates that the abstraction has introduced a spurious counterexample, requiring us to revise this abstraction and revert to the original statement.\looseness = -1

 \begin{algorithm}[b!] 
{
\scriptsize             
\setstretch{0.65}            
 	\caption{The R-CEGAR process} 
 	\label{alg1} 
 	\DontPrintSemicolon
 		\KwInput{$\hat{M}$: the abstract model generated by our method, $P$: the target property to prove;}
 		\KwOutput{$res \in \{\mathit{SAFE}, \mathit{UNSAFE}\}$;}
 		\While{True} {
 		$cex \gets \text{PropertyChecking}(\hat{M}, P);$ \label{cexcheck}  \;
			\eIf{$cex = \emptyset$}
            {%
                    \tcp{The property is proved.}%
                    \Return \text{SAFE} }{
                    $cex\_generalized \gets \text{Generalize}(cex)$ \;\label{cexgen}
                    $invalid\_statements  \gets \text{Check}(cex\_generalized)$ \; \label{detect}
                    \eIf{$invalid\_statements  = \emptyset$}{
                        \Return \text{UNSAFE}\;\label{check-result}
                    }{
                        $regions \gets Locate(invalid\_statements)$\;\label{locate}
                        $\hat{M} \gets 
                        \text{Refine}(\hat{M}, invalid\_statements, regions)$\;\label{refine}
                    }
                    }

}
}

 \end{algorithm}

The overall R‑CEGAR process is illustrated in Algorithm~\ref{alg1}. The algorithm takes as input the initial abstract state transition system $\hat{M}$ and the target property $P$. In each iteration, the algorithm performs property checking (line~\ref{cexcheck}). If a violation is detected, it extracts a counterexample and applies the reduction technique to generalize the trace (line~\ref{cexgen}).
It then checks if any abstract statements contribute to the property violation (line~\ref{detect}). If such statements are detected, we first locate them through the AST-based symbolic representation (line~\ref{locate}), mapping each invalid abstract statement back to its corresponding RTL region and rewrite point, and then revise them accordingly (line~\ref{refine}); otherwise, the property is indeed violated, implying that the original model also fails under this property (line~\ref{check-result}).\looseness=-1

\vspace{-2mm}
\section{Experiment}
\label{sec::exp}
In this section, we aim to answer the following questions through a comparative evaluation of NeuroAbs: \textbf{(RQ1)} What is the performance of NeuroAbs in generating the abstract statements? 
\textbf{(RQ2)} Can the abstraction accelerates model checking algorithms in the construction of formal proofs? 
 \textbf{(RQ3)} When model checking algorithms can not conclude with a formal proof, can our abstraction method help to explore more state transitions for bug-finding? 


\begin{table}[t]
\centering
\renewcommand{\arraystretch}{0.83} 
\setlength{\tabcolsep}{9pt} 
\caption{The statistics of benchmark designs in the experiment.}
\vspace{-3mm}
\label{Tab::benchmark}
\scalebox{0.75}{
\setlength{\tabcolsep}{1mm}
{
\begin{tabular}{|c|c|c|c|c|c|}
\hline
\begin{tabular}[c]{@{}c@{}}Source\\ 
Benchmarks\end{tabular} & \begin{tabular}[c]{@{}c@{}}Verification\\ Task\end{tabular}                       & \begin{tabular}[c]{@{}c@{}}Number of \\ Scenarios\end{tabular} & \begin{tabular}[c]{@{}c@{}}\# RTL\\ Lines\end{tabular} & \begin{tabular}[c]{@{}c@{}}\# Input\\ Vars\end{tabular} & \begin{tabular}[c]{@{}c@{}}\# State\\ Vars\end{tabular} \\ \hline \hline
\multirow{2}{*}{PicoRV32}                                   & \multirow{2}{*}{\begin{tabular}[c]{@{}c@{}}Generic\\ Inst.-specific\end{tabular}} & \multirow{2}{*}{8}                                             & \multirow{2}{*}{3079}                                  & \multirow{2}{*}{82}                                     & \multirow{2}{*}{209}                                    \\  
                                                            &                                                                                   &                                                                &                                                        &                                                         &                                                         \\ \hline
\multirow{2}{*}{Piccolo}                                    & \multirow{2}{*}{\begin{tabular}[c]{@{}c@{}}Generic\\ Inst.-specific\end{tabular}} & \multirow{2}{*}{22}                                            & \multirow{2}{*}{14943}                                 & \multirow{2}{*}{219}                                    & \multirow{2}{*}{381}                                    \\  
                                                            &                                                                                   &                                                                &                                                        &                                                         &                                                         \\ \hline
\multirow{2}{*}{Flute}                                      & \multirow{2}{*}{\begin{tabular}[c]{@{}c@{}}Generic\\ Inst.-specific\end{tabular}} & \multirow{2}{*}{22}                                            & \multirow{2}{*}{17228}                                 & \multirow{2}{*}{279}                                    & \multirow{2}{*}{407}                                    \\ 
                                                            &                                                                                   &                                                                &                                                        &                                                         &                              \\ \hline                         

\multirow{2}{*}{I2C}                                      & \multirow{2}{*}{\begin{tabular}[c]{@{}c@{}}Reg-interface\end{tabular}} & \multirow{2}{*}{3}                                            & \multirow{2}{*}{1257}                                 & \multirow{2}{*}{157}                                    & \multirow{2}{*}{47}                                    \\ 
                                                            &                                                                                   &                                                                &                                                        &                                                         &                              \\ \hline  
                                                        \end{tabular}
}
}
\vspace{-4mm}
\end{table}

\subsection{Experiment Setup}
The experiments are conducted on a machine with a 2.9~GHz Intel Xeon Platinum 8375C CPU and 256~GB RAM. We adapt the open-source RTL synthesis tool Yosys to perform RTL static analysis and use Pyverilog to tranform the RTL code to AST. For the LLM, we utilize the gpt-4o-mini-2024-07-18's API to perform the abstraction since it balances the cost and the quality of the generated result. For few-shot learning, we provide 8 input-output examples for NeuroAbs to learn how to perform abstraction. We translate the rewritten RTL statements into symbolic formulas and use the SMT solver Z3 to check whether each generated abstraction satisfies the required over-approximation constraint.

We integrate our abstraction method with two model checkers: RIC3~\cite{su2024predicting} and AVR~\cite{goel2020avr}. AVR was the champion of the 2020's hardware model checking competition (HWMCC)~\cite{preiner2020hardware}. It incorporates various word-level model checking algorithms and supports parallel execution. We utilize its development branch for our experiments. RIC3 won the championship at the 2024's HWMCC~\cite{biere2024hardware}. It includes several bit-level model checking algorithms running in parallel. We used the version prepared for the 2024's competition in our experiments. Note that both AVR and RIC3 integrate a variety of advanced abstraction techniques and have demonstrated highly competitive performance in HWMCC, making them substantially stronger baselines than prior abstraction approaches such as IC3IA. Our goal is to demonstrate that NeuroAbs can further improve the efficiency of these model checkers, even when they are already equipped with traditional abstraction mechanisms.

\subsection{Benchmarks}

For our experiments, we need benchmark designs that come with: 1) the RTL implementation, 2) a design specification, and 3) a set of formal properties to be verified. However, many existing open-source RTL designs lack a good design specification in the first place and few are equipped with formal properties. 
Therefore, we turn to the verification tasks of open-source RISC-V processors, as RISC-V processor designs are based on well-defined standards, and there are existing works providing formal properties to verify.
The verification tasks we use are over three processor cores, PicoRV32~\cite{PICO},  Piccolo~\cite{Piccolo} and Flute~\cite{Flute}. PicoRV32 is verified using the RISCV-formal framework~\cite{waterman2014risc}, while Piccolo~\cite{Piccolo} and Flute~\cite{Flute} are verified using the instruction-level abstraction (ILA) refinement checking~\cite{huang2018instruction}. To demonstrate applicability beyond processors, we also include an I2C peripheral, verifying properties such as read-after-write consistency in its register interface. Table~\ref{Tab::benchmark} summarizes the statistics of our benchmarks, which shows that the benchmarks used in experiments are non-trivial in terms of RTL code size, with a total of 55 verification scenarios identified across three designs. 
Among these verification scenarios, some check the correctness of a single instruction or an instruction class (instruction-specific), while others check generic properties of the RTL, the sanity of the specified properties, or register-interface consistency.
\looseness = -1

\begin{table}[t]
\centering
\caption{Ablation study of NeuroAbs}
\vspace{-3mm}
\label{tab::result_abstraction}
\scalebox{0.72}{
\setlength{\tabcolsep}{0.80mm}
{

\begin{tabular}{l|cccc}
                                \specialrule{0em}{1.5pt}{1.5pt}
\toprule[1.2pt]
\specialrule{0em}{1.5pt}{1.5pt}
                               
                                & \begin{tabular}[c]{@{}c@{}}OA \\ rate\end{tabular} & Degradation & \begin{tabular}[c]{@{}c@{}}\# avg. strict \\ OA  \end{tabular} & Degradation  \\
                                \specialrule{0em}{1.5pt}{1.5pt}
\hline
\specialrule{0em}{1.5pt}{1.5pt}
NeuroAbs                        & 95.27\%                                                  & -           & 236.17                                                            & -           \\
w.o. few-shot                       & 83.51\%                                                  & 12.34\%     & 136.00                                                            & 42.41\%     \\
w.o. few-shot \& custom inst. & 96.46\%                                                  & -1.25\%     & 3.83                                                              & 98.38\%    
\\ 
\specialrule{0em}{1.5pt}{1.5pt}
\toprule[1.2pt]
\specialrule{0em}{1.5pt}{1.5pt}
\end{tabular}
}
\vspace{-3mm}
}
\begin{tablenotes} \fontsize{6}{6}\selectfont
\item $^\star$ A rewritten statement is an over-approximation (OA) as long as it complies with \cref{eqa::overapp}. This does not rule out equivalent statements.
\item $^\dagger$ This column reports the averaged number of strict over-approximated statements, which must comply with both \cref{eqa::overapp} and \cref{eqa::uneq}.
\end{tablenotes} 
\vspace{-5mm}
\end{table}

\begin{table}[h]
\centering
\caption{Average unrolling depths explored by BMC w./w.o. NeuroAbs.}
\vspace{-3mm}
\label{tab::bound}
\footnotesize

\renewcommand{\arraystretch}{1.12} 
\scalebox{0.75}{
\setlength{\tabcolsep}{0.88mm}
{

\begin{tabular}{@{}c  c  c c c c@{}}
\toprule
Design$^*$ & Runtime (s) &
\cellcolor[HTML]{C5D9F1}With NeuroAbs &
\cellcolor[RGB]{225,225,226}Baseline$^\dagger$ &
\makecell{Time to Reach\\ Baseline Max Depth (s)} &
Improvement (\%) \\
\midrule
Piccolo & \multirow{2}{*}{23600} & \textbf{125.36} & 109.36 & 15482.67 & 34.40\% \\
Flute   &                        & \textbf{225.09} & 129.00 &  5668.58 & 75.98\% \\
\bottomrule
\end{tabular}
}
}
\begin{tablenotes} \fontsize{6}{6}\selectfont
\item $^\dagger$ Baseline refers to the results without NeuroAbs.
\item $^*$ Since all PicoRV32-related and I2C-related cases have been formally proven in the previous experiment, we do not include them in the BMC experiments.
\end{tablenotes}
\vspace{-5mm}
\end{table}

\begin{table*}[t]
\centering
\caption{Experiment results on representative cases for BMC with or without NeuroAbs.}
\vspace{-3mm}
\label{tab::detail_bmc}
\renewcommand{\arraystretch}{1} 
\scalebox{0.74}{
\setlength{\tabcolsep}{0.85mm}
{

\begin{tabular}{|c|c|c|c|c|c|c|c|c|c|c|}
\hline
Benchmark                     & \begin{tabular}[c]{@{}c@{}}Verification \\ Task\end{tabular} &  \begin{tabular}[c]{@{}c@{}}\# Variable Index \\ (w./w.o. Abstraction)\end{tabular} & \begin{tabular}[c]{@{}c@{}}Total LLM \\ time  (s) $^*$ \end{tabular} & \begin{tabular}[c]{@{}c@{}}Avg. LLM \\ time / stmt (s)\end{tabular} & \begin{tabular}[c]{@{}c@{}}CEGAR \\ time (s)\end{tabular} & \begin{tabular}[c]{@{}c@{}}\# CEGAR \\ Iter\end{tabular} & \begin{tabular}[c]{@{}c@{}} \cellcolor[HTML]{C5D9F1}\# Bound with \\ \cellcolor[HTML]{C5D9F1}NeuroAbs \end{tabular} & \begin{tabular}[c]{@{}c@{}}{\cellcolor[RGB]{225, 225, 226}\# Bound}\\ {\cellcolor[RGB]{225, 225, 226}Baseline}\end{tabular} & \begin{tabular}[c]{@{}c@{}}Time to Reach \\ Baseline Max Bound   (s)\end{tabular} & Improvement (s)  \\ \hline \hline
\multirow{4}{*}{Piccolo} 
                         & OR-ILA                                                &\textbf{118188}/129316 & 361.79                                                         & 4.41                                                            & 2.88                                                      & 2                                                        & \textbf{99}                                                                     & 87                                                          & 15378.68                                                                          & 34.84\%         \\
                         & ANDI-ILA                                     & \textbf{116128}/119628               & 275.80                                                         & 3.40                                                            & 4.05                                                      & 3                                                        & \textbf{123}                                                                    & 106                                                         & 14126.45                                                                          & 40.14\%         \\
                         & JAL-ILA                                               & \textbf{111567}/116969      & 305.72                                                         & 3.40                                                            & 0.62                                                      & 1                                                        & \textbf{190}                                                                    & 163                                                         & 13982.43                                                                          & 40.75\%         \\
                         & LUI-ILA   & \textbf{111546}/115882                                                  & 274.46                                                         & 3.19                                                            & 5.79                                                      & 3                                                        & \textbf{200}                                                                    & 180                                                         & 15396.60                                                                          & 34.76\%         \\ \hline
\multirow{4}{*}{Flute}   
& SUB-ILA                                                    &\textbf{299217}/1273706 & 632.07                                                         & 2.13                                                            & 6.83                                                      & 1                                                        & \textbf{221}                                                                    & 138                                                         & 6978.98                                                                           & 70.43\%         \\
                         & XORI-ILA                                                & \textbf{295565}/1272600    & 611.42                                                         & 2.09                                                            & 13.77                                                     & 1                                                        & \textbf{224}                                                                    & 133                                                         & 5351.54                                                                           & 77.32\%         \\
                         & BLTU-sanity                                  & \textbf{306462}/1278074               & 663.89                                                         & 2.31                                                            & 42.00                                                     & 2                                                        & \textbf{216}                                                                    & 123                                                         & 5252.36                                                                           & 77.74\%         \\
                         & BGE-sanity                                    & \textbf{300650}/1275355               & 604.23                                                         & 2.04                                                            & 14.26                                                     & 2                                                        & \textbf{220}                                                                    & 129                                                         & 5552.50                                                                           & 76.47\%         \\
                         \hline   
\end{tabular}
}
}

\begin{tablenotes}\fontsize{7}{7}\selectfont
\item $^\dagger$ Due to page limits, we are unable to include all detailed results in this table. Additional results 
are available at: [\href{https://doi.org/10.6084/m9.figshare.30633074}{\textbf{Link}}]
\item $^*$ The reported time in this column includes the time spent on LLM inference, SMT solving, and AST construction and traversal.
\end{tablenotes} 
\vspace{-3mm}
\end{table*}

\subsection{Effectiveness of our LLM in Performing Abstraction}
\begin{sloppypar}

To address \textbf{RQ1}, we conduct an ablation study to evaluate the contribution of individual components in the NeuroAbs framework for abstraction construction. Specifically, we compare the full NeuroAbs against degraded variants in which certain design elements are removed step by step. First, we disable few-shot learning, thereby removing example-based guidance that helps the LLM produce correct abstract statements. Then, we remove the custom instruction prompt, leaving the model without sufficient contextual understanding of the abstraction task. This configuration corresponds to the default behavior of gpt-4o-mini-2024-07-18.\looseness = -1

We evaluate performance using two metrics. The first is the over-approximation rate (OA rate), which measures the percentage of generated statements that correctly produce an over-approximate version, as validated by the condition in \cref{eqa::overapp}. The second metric is the average number of strict over-approximations. In addition to satisfying the over-approximation condition, this metric requires that the abstracted statement be semantically different from the original, such that it is not trivially equivalent to the source statement. Formally, the abstract statement must satisfy both the condition in \cref{eqa::overapp} and the following criterion:


 \begin{equation}
 \label{eqa::uneq}
\begin{split}
\forall V,\ \exists Y,\ \left[ S(V, Y) \neq  \hat{S}(V, Y) \right]
\end{split}
\end{equation}
\end{sloppypar}

\Cref{tab::result_abstraction} presents the overall results of the ablation study. From this table, two key observations can be made. First, removing the few-shot learning mechanism significantly reduces the model’s ability to generate correct over-approximated statements, as it lacks concrete guidance on how to construct valid abstractions. This leads to a decreased OA rate (83.51\%). Second, when custom instructions are also removed, the OA rate appears to increase slightly to 96.46\%. However, this improvement is misleading. Without the custom instructions, NeuroAbs loses awareness of the objective of making an abstraction and instead tends to greedily reproduce the original statements. As a result, the number of strict over-approximation drops drastically, with an average of only 3.83 statements---representing a 98.38\% decrease compared to the fully configured NeuroAbs. Overall, these results emphasize the importance of both few-shot learning and custom instruction prompts: together, they enable NeuroAbs to produce meaningful and diverse abstractions.

\begin{figure}[t]
    \centering
    \includegraphics[width=0.9\linewidth]{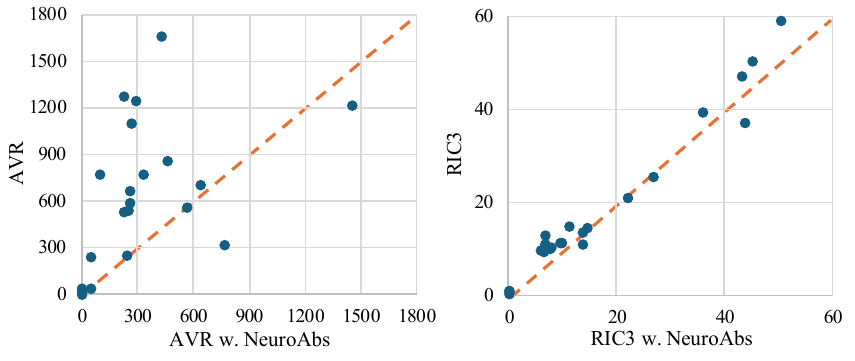}
                        \vspace{-3mm}
                        \caption{The runtime (measured in seconds) comparison of reaching a formal proof with or without NeuroAbs. The mean values are: AVR = 607.03, AVR w. NeuroAbs = 334.56; 
    RIC3 = 18.71, RIC3 w. NeuroAbs = 16.94.}
    \label{fig:exp_formal}
\vspace{-5mm}
\end{figure}

\subsection{Effectiveness of Abstraction in Concluding Formal Proofs}

To address RQ2, we evaluate the effectiveness of our abstraction technique in facilitating formal proofs. \cref{fig:exp_formal} compares the verification runtime with and without NeuroAbs across different model checkers, with a 3600-second timeout for each case. When integrated with AVR, NeuroAbs significantly reduces proof time, achieving an overall runtime improvement of 44.89\%. Although AVR already includes advanced abstraction mechanisms such as syntax-guided abstraction and uninterpreted function abstraction, NeuroAbs further complements them and enhances performance. For RIC3, integration also improves runtime, though the gains are marginal as RIC3 is already running fast for these cases. \looseness = -1

\begin{figure*}[ht]
    \centering

    \includegraphics[width=0.9\linewidth]{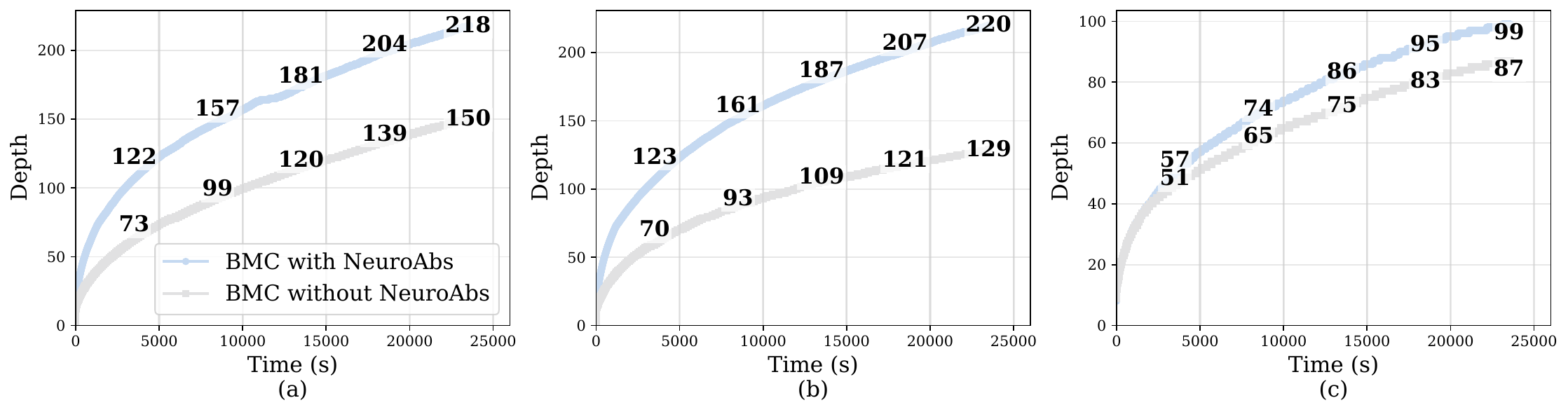}
            \vspace{-3mm}
            \caption{Comparison of Depth vs. Time relationship between BMC with and without NeuroAbs across three representative examples.}
    \label{fig:exp_bound}
\vspace{-3mm}
\end{figure*}

\subsection{Effectiveness of Abstraction in BMC}
For cases where a formal proof cannot be concluded in reasonable time, we analyze the effectiveness of our abstraction method in accelerating the BMC algorithm to address \textbf{RQ3}. Although these cases do not yield a formal proof, faster BMC runs allow deeper exploration within the same runtime, improving the chance of finding bugs or gaining confidence in correctness. Since AVR is not as competitive as RIC3, as shown by the prior result, this experiment solely utilizes the BMC implementation in RIC3 for comparison. \cref{tab::bound} presents the average achieved depths with and without NeuroAbs. 
We allow BMC to run for up to 23600 seconds per case to capture meaningful differences in the depths reached by each configuration. In this table, the column ``Time to Reach Baseline Max Depth'' indicates the time required for BMC with NeuroAbs to reach the maximum cycles achieved by the baseline.\looseness=-1


As shown in \cref{tab::bound}, integrating NeuroAbs into the BMC algorithm substantially enhances its efficiency, enabling deeper exploration within the same runtime. For example, in the ``Piccolo'' case, the average depth increases from 109.36 to 125.36. It also greatly reduces the time required to reach the baseline’s maximum depth; in the ``Flute'' case, the runtime decreases by 75.98\%, from 23,600 seconds to 5,668.58 seconds. Detailed results for representative cases are also provided in \cref{tab::detail_bmc}. Overall, NeuroAbs consistently improves BMC performance across the given tasks. \looseness=-1


\cref{tab::detail_bmc} also reports the total runtime, the number of refinement iterations, the cumulative LLM runtime, and the average time required for the LLM to abstract each statement. The results show that the initial abstract model can be refined efficiently, requiring only a small amount of time and few iterations. On average, the refinement process takes 7.26 seconds and 1.65 iterations. This efficiency stems from the high accuracy of NeuroAbs, which produces high‑quality abstractions that greatly reduce the refinement effort.

Furthermore, we plot the wall-clock time against the number of reached bounds for three representative cases, as shown in \cref{fig:exp_bound}. From this figure, we observe that as the time increases, the gap between the depths reached by BMC with and without NeuroAbs continues to widen. For example, in the middle graph, at 10,000 seconds, the depth difference is 68 (161 vs. 93), but by 20,000 seconds, this gap increases to 86 (207 vs. 121).  This trend demonstrates the effectiveness of the abstraction in simplifying the circuit structure, which reduces the complexity of the unrolled expressions in BMC. As a result, the solving process becomes increasingly efficient over time, particularly at larger unrolling depths.

Regarding the inference time for each statement, the LLM demonstrates efficient abstraction performance, requiring on average 3.70 seconds to rewrite a statement. The total LLM runtime, however, is relatively longer because abstractions are generated sequentially rather than in parallel. Nevertheless, even when accounting for both LLM runtime and CEGAR refinement time, our method still outperforms the baseline. Further speedups could be achieved by parallelizing the rewriting process to avoid linear growth in total runtime. \looseness = -1

\subsection{Discussion}

To understand why NeuroAbs improves the efficiency of model checkers, we compare hardware sizes with and without our abstraction technique for the same representative cases shown in \cref{tab::detail_bmc}. Specifically, we translate the RTL design into an And-Inverter-Gate (AIG) circuit~\cite{biere2007aiger} using Yosys~\cite{wolf2016yosys}, and record the maximum variable index, as summarized in \cref{tab::detail_bmc}. In the AIG format, the maximum variable index equals the sum of inputs, latches, and AND gates. The results show that the maximum variable index are greatly reduced. This reduction comes from the way our abstraction simplifies large parts of the original logic, thereby lowering circuit complexity and accelerating the model checking algorithm.  \looseness=-1

\begin{table}[t]

\caption{Common RTL Abstraction Types.}
\vspace{-3mm}
\label{tab::abstraction type}
\renewcommand{\arraystretch}{1} 
\scalebox{0.77}{
\setlength{\tabcolsep}{1mm}
{

\begin{tabular}{|c|l|l|}
 \hline
\textbf{Type} & \multicolumn{1}{c|}{Before NeuroAbs}                                                                                                                                      & \multicolumn{1}{c|}{After NeuroAbs}                                                                                                                \\ \hline \hline
Unary         & c = $\sim$a;                                                                                                                                         & c = 'bx;                                                                                                                      \\ \hline
Binary        & out = a \& b;                                                                                                                                        & out = 'bx;                                                                                                                    \\ \hline
Ternary       & out = (PIPELINED) ?  a+ b : a - b;                                                                                                                   & out = (PIPELINED) ?  a+ b : 'bx;                                                                                              \\ \hline
Case          & \begin{tabular}[c]{@{}l@{}}\textbf{case} (inst) \textbf{begin} \\ 1'b1: result = a || b; \\ 1'b0: result = a*b; \textbf{end}\end{tabular}                                       & \begin{tabular}[c]{@{}l@{}}\textbf{case} (inst) \textbf{begin}  \\ 1'b1: result = a || b; \\ 1'b0: result = 'bx; \textbf{end}\end{tabular}                \\ \hline

For       & \begin{tabular}[c]{@{}l@{}}\textbf{for} (i = 0; i \textless 32; i = i+1)  \\ regs = 0;\end{tabular}          
& \begin{tabular}[c]{@{}l@{}}\textbf{for} (i = 0; i \textless 32; i = i+1)  \\ regs = 'bx;\end{tabular}                                  \\ 
\hline
Always   & \begin{tabular}[c]{@{}l@{}}\textbf{always} @(\textbf{posedge} clk) \\ \textbf{begin}  compare \textless{}= |\{beq, bne\}; \\ ari \textless{}= |\{add,   sub\} ; \textbf{end}\end{tabular} & \begin{tabular}[c]{@{}l@{}}\textbf{always} @(\textbf{posedge} clk) \\ \textbf{begin}  compare \textless{}= |\{beq, bne\}; \\ ari \textless{}= 'bx ; \textbf{end}\end{tabular} \\ 
\hline
Initial       & \begin{tabular}[c]{@{}l@{}}\textbf{initial} \textbf{begin}  \\ receive \textless{}= 'b0 ; send \textless{}= 'b0 ; \textbf{end}\end{tabular}                                      & \begin{tabular}[c]{@{}l@{}}\textbf{initial} \textbf{begin}  \\ receive \textless{}= 'bx ; send \textless{}= 'bx ; \textbf{end}\end{tabular}    
\\ \hline
\end{tabular}
}
}
\vspace{-5mm}
\end{table}

\subsection{Supported Types of RTL Constructs}
We further summarize the types of RTL constructs that NeuroAbs can abstract based on the observations. As shown in \cref{tab::abstraction type}, NeuroAbs supports several common abstraction types in RTL designs. It successfully abstracts statements with unary and binary operators from non-blocking assignments and entire always blocks containing multiple blocking statements. It also flexibly abstracts branch structures, such as ternary operators and case statements, according to the given property. These results show that NeuroAbs
effectively handles diverse abstraction scenarios. \looseness=-1

\section{Conclusion}
\label{sec::conclusion}
In this paper, we introduce NeuroAbs, a neuro-symbolic abstraction framework that leverages LLMs to achieve flexible and automated abstraction for RTL models. The framework starts with LLM-assisted RTL analysis to identify suitable signals and statements for abstraction, followed by AST-guided LLM-based abstraction to generate an initial abstract model. Finally, SMT checking and the CEGAR process are employed to maintain soundness and iteratively refine the model when necessary. Experimental results demonstrate that NeuroAbs can effectively improve the efficiency of formal property verification on various RTL designs.\looseness=-1


\begin{acks}
This work is supported by the National Natural Science Foundation of China (grant no. 62304194).
\end{acks}






\balance
\bibliographystyle{ACM-Reference-Format}
\bibliography{sample-base}
\end{document}